\documentclass[12pt]{article}
\begin{document}
\begin{center}
\baselineskip=35pt
 {\LARGE \bf Coherence effects in the transition
radiation  spectrum and
 practical consequences\footnote{Submitted to Nuclear Instruments and %
 Methods in Physics Research Section A}}
 \end{center}
\vspace{.9cm}
\begin{center}
\baselineskip=20pt
 {Gian Luca Orlandi\footnote{Tel.:
+39-0649903609; Fax: +39-0649387075;\\  E-mail
Gianluca.Orlandi@iss.infn.it } \\  Istituto Superiore di Sanit\`{a}, %
Physics Laboratory, and %
 Istituto Nazionale di Fisica Nucleare, Gruppo Sanit\`{a} Roma1
 \\  Viale Regina Elena 299, 00161 Roma, Italy }
\end{center}
\vspace{.9cm}
\begin{abstract}
 \baselineskip=16pt
In the framework of the {\em pseudophotons method\/} and under the
hypothesis of the {\em far field approximation\/} the formal
evaluation of the transition radiation spectrum produced by a
three-dimensional charged beam interacting with a metallic target
is carried out. When spatial incoherent conditions are occurring,
the transverse size of a three-dimensional beam affects the
spectrum via a suitable dependence on the Fourier transform of the
transverse distribution, while the temporal coherent enhance of
the spectrum depends only on the longitudinal beam size via the
{\em longitudinal form factor\/} in the same manner as in the case
of a one-dimensional beam. Spatial incoherent effects, arising at
short wave-length in the spectrum, on the one hand allow a better
comprehension of some aspects of the transition radiation
phenomelogy - e.g. an analytical formula for the total radiated
energy - and on the other hand offer new perspectives in the
practical application of the transition radiation as diagnostics
method of the transverse beam size in a particle accelerator.
\end{abstract}
 \thispagestyle{empty}

\newpage
 \baselineskip=18pt
% main text
\section{Introduction}
%\label{}

Transition radiation theory describes those radiative phenomena
produced by high energy charged particles crossing the interface
between two media with different dielectric properties. Typical
transition radiation events are represented by relativistic
charged particles crossing in the vacuum a thin metallic layer. As
natural extension, such a category of events also includes the
diffraction radiation phenomena, i.e., the radiative mechanisms
observed when a charged beam passes through a circular or
rectangular slit over a metallic screen.

In the past few years the transition radiation theory, based on
the {\em single particle\/} formalism describing a point-like
charge colliding with an ideal metallic target of infinite size
\cite{termik,gari,bass,ginz}, attracted much interest in those
aspects concerning the diffractive modifications affecting such
formalism as the finite size of the screen is considered
\cite{barry,shulga,poty,ccov}. The aim of the present paper is to
contribute to an extension of the transition radiation theory by
including in the formalism the {\em source size\/} effects arising
as the charged beam can be realistically described by a
three-dimensional probability distribution.

The {\em pseudophotons method\/}, describing the transition
radiation produced by a high energy beam as a radiation field
scattered off a metallic screen, is a powerful formalism
\cite{termik,jack} allowing to calculate the diffractive
modifications in the spectrum due to the finite size of the
metallic target and/or to the presence of a slit on it. In the
framework of such a formalism and supposing - as reasonable in
practical situations - the linear size of the detector surface
much smaller than the distance between the scattering surface and
the detector itself ({\em far field approximation\/}), the
complete formal expression of the transition radiation spectrum
produced by a three-dimensional electron bunch colliding with a
metallic target is obtained in this paper. As different from the
classical articles \cite{schiff,nodsax,hirsch}, which are
dedicated to the collective effects observable at long wave-length
in the synchrotron radiation spectrum emitted by one-dimensional
electron beam with zero transverse dimension moving on a circular
orbit, the present paper treats transition radiative phenomena
concerning three-dimensional charged beam and, consequently,
derives physical consequences.

\section{Transition radiation spectrum produced by %
a three-dimensional charged beam in the framework of the
pseudophotons method}

It has been shown \cite{nodsax,hirsch} that the synchrotron
radiation spectrum emitted by $N$ electrons, collected in a
longitudinally extended bunch with zero transverse dimension
moving on a circular orbit, is described by the following
expression
\begin{equation}
\frac{d I(\omega)}{d \omega}=\frac{d I_{e}(\omega)}{d \omega}[N%
+N(N-1)F(\omega)]\label{coe},
\end{equation}
where $\frac{d I_{e}(\omega)}{d\omega}$ - the intensity per
frequency unit - is related to the radiative mechanism produced by
the single electron in the bunch ({\em single particle
spectrum\/}). The {\em longitudinal form factor\/} $F(\omega)$ can
be stated - in the limit of a continuous charge distribution - as
the squared module of the Fourier transform of the longitudinal
probability distribution function $\rho_{z}(z)$:
\begin{equation}
F(\omega)=\left| \int\limits_{-\infty}^{+\infty} d z\,\rho_{z}(z)%
e^{i(\omega z/c)} \right|^2\label{ff}.
\end{equation}
Due to its relativistic origin, Eq.(\ref{coe}) continues to
maintain its validity for one-dimensional beams if, instead of
considering the synchrotron radiative mechanism, we consider the
transition one. Then, we assume that the $N$ electrons - collected
in a three-dimensional bunch  and in linear uniform motion with
common velocity $\vec{w}$ - hit normally a flat metallic screen
producing forward and backward transition radiation. Whether the
procedure of substituting in (\ref{coe}) the {\em longitudinal
form factor\/} (\ref{ff}) with a three-dimensional one
\begin{equation}
F(\vec{k})=\left|\int d \vec{r}\,\rho(\vec{r})%
e^{i\vec{k}\cdot\vec{r}}\right|^2\label{triff}
\end{equation}
is the correct one when the charged beam is described by a
three-dimensional continuous probability distribution function
$\rho(\vec{r})$, represents the subject of the following part of
this paper.

The frequency distribution of the transition radiation intensity
can be obtained describing the radiative mechanism in terms of the
{\em pseudophotons method\/} \cite{termik,jack}, which allows to
directly include in the calculations also screen size effects.
According to the {\em Huygens-Fresnel principle\/}, the transition
radiation field can be therefore imagined as the diffractive
propagation of a {\em pseudophotons\/} field scattered off the
metallic screen. In practice the harmonic components of the
transition radiation field, meant as scattered components of the
transverse electromagnetic (e.m.) field moving with the charge,
can be analytically expressed via the {\em integral theorem of
Helmholtz and Kirchhoff\/} \cite{bw}. Such integrals, under the
assumption of the {\em far field approximation\/}, can be simply
calculated \cite{termik,casver} by Fourier transforming the {\em
pseudophotons\/} field components with respect to the coordinates
$(x,y)$ of the metallic target surface $S$. Therefore, at large
distance $R$ from the target the transition radiation electric
field components $(E_x^{tr}, E_y^{tr})$, expressed in terms of the
corresponding components $(E_x^{ps}, E_y^{ps})$ of the scattered
{\em pseudophotons\/} field, are:
\begin{eqnarray}
E_{x,y}^{tr}(k_x,k_y,\omega)=&&\frac{k}{2 \pi R}\times\nonumber\\
&&\times\int\limits_S d xd y\,E_{x,y}^{ps}(x,y,\omega)%
e^{-i(k_xx+k_yy)},\nonumber\\ \label{trans}
\end{eqnarray}
where $k=\omega/v=\sqrt{k_x^2+k_y^2+k_z^2}$ is the module of the
transition radiation wave-vector in a non-dispersive medium with
phase-velocity $v=c/\sqrt{\epsilon\mu}$.

\subsection{Charge electromagnetic field as pseudophotons field}

Before expressing the {\em pseudophotons\/} field ($\vec{E}^{ps}$)
in terms of the the e.m. field ($\vec{E}^{ch}$) travelling with
the electron bunch, it is better to make clear the scenario.

It is assumed that a charged beam, moving along the $z$ axis with
constant velocity $\vec{w}=(0,0,w)$, hits a flat surface $S$, made
of an ideal conductor (dielectric constant
$\epsilon_{cond}=\infty$) and lying on the $(x,y,z=0)$ plane of
the reference frame. A non-dispersive medium with permeability
$\mu=1$ and characteristic phase velocity $v=c/\sqrt{\epsilon}$
surrounds the surface. Furthermore it is assumed that $N$
electrons, forming a three-dimensional bunch at rest with respect
to the center-of-mass frame, are described by a distribution
function $\rho(\vec{r})$ ``frozen'' in time:
\begin{eqnarray}
\rho(\vec{r},t)=\rho(\vec{r},0)%
=\int \frac{d\vec{k}}{(2\pi)^3}\,e^{i\vec{k}\cdot%
\vec{r}}\rho(\vec{k})\label{con-sp-dis}.
\end{eqnarray}
This assumption, in accordance with the observation that the
radiative process at a given time only depends on those electrons
in the bunch hitting at the same time the metallic screen,
corresponds to treat the $N$ electrons as uncorrelated and
described by a probability distribution function and a
corresponding Fourier transform expressed as:
$\rho(\vec{r})=\rho_x(x)\rho_y(y)\rho_z(z)$ and
$\rho(\vec{k})=\rho_x(k_x)\rho_y(k_y)\rho_z(k_z)$ respectively.

It can be demonstrated that the Fourier transform of the electric
field moving with the charge is
\begin{eqnarray}
\vec{E}^{ch}(\vec{k},\omega)=&&(8\pi^2 Ne)\times\nonumber\\
&&\times\frac{[-i(\vec{k}/\epsilon)+i(\omega\vec{w}/c^2)]}%
{[k_x^2+k_y^2+k_z^2-(\omega/v)^2]}\rho(\vec{k})%
\delta(\omega-\vec{k}\cdot\vec{w}),\nonumber\\ \label{elec}
\end{eqnarray}
The longitudinal field component $E_z$, proportional to the
reciprocal of the square of the relativistic Lorentz factor
$\gamma=E/mc^2={(1-(w/v)^2)^{-1/2}}$, can be neglected with
respect to the electric field components $E_{x,y}$ transverse to
the charge velocity $\vec{w}$.

Therefore, the harmonic components of the transverse electric
field traveling with the charge follows from Eq.(\ref{elec}):
\begin{eqnarray}
E_{x,y}^{ch}(\vec{r},\omega)=-&&\frac{ie^{i(\omega
z/w)}Ne\rho_z(\omega/w)}{\pi w}\times\nonumber\\
&&\times\int d k_x d k_y\,\frac{k_{x,y}\rho_x(k_x)\rho_y(k_y)%
e^{i(k_xx+k_yy)}}%
{\epsilon[k_x^2+k_y^2+\alpha^2]},\nonumber\\ \label{carica}
\end{eqnarray}
where $\alpha=\omega/(w\gamma)$. The {\em pseudophotons\/} field
can be finally derived from the constraint of the boundary
conditions over the metallic surface imposed on the transverse
component of the total field:
\begin{equation}
E_{x,y}^{ps}(x,y,z=0)+E_{x,y}^{ch}(x,y,z=0)=0\label{bound}.
\end{equation}
Then from Eq.(\ref{bound}) the explicit expression of the harmonic
components of the {\em pseudophotons\/} field is:
\begin{eqnarray}
E_{x,y}^{ps}(\vec{r},\omega)=%
&&\frac{i e^{-i(\omega z/c)}Ne\rho_z(\omega/w)}{\pi
w}\times\nonumber\\
&&\times\int d k_x d k_y\,%
\frac{k_{x,y}\rho_x(k_x)\rho_y(k_y)e^{i(k_xx+k_yy)}}%
{\epsilon[k_x^2+k_y^2+\alpha^2]}.\nonumber\\ \label{esplipse}
\end{eqnarray}

\subsection{Transition radiation field as scattered pseudophotons %
field}

With respect to a spherical $(r,\theta,\phi)$ coordinates
reference and assuming to observe the radiation far enough from
the metallic screen along the polar $z$ axis $(z=R\cos \theta
\approx R)$, the harmonic components of the transition field
Eq.(\ref{trans}) can be explicitly expressed via
Eq.(\ref{esplipse}) as
\begin{eqnarray}
E_{x,y}^{tr}(\vec{\tau},\omega)=&&\frac{ikNe%
\rho_z(\omega/w)e^{-i(\omega R/c)}}{2\pi^2Rw}\times\nonumber\\
&&\times\int\limits_{S}d \vec{\xi}\,\int%
d\vec{\kappa}\,\frac{\kappa_{x,y}\rho_x(\kappa_x)\rho_y(\kappa_y)%
e^{-i(\vec{\tau}-\vec{\kappa})\cdot\vec{\xi}}}%
{[\kappa^2+\alpha^2]}\nonumber\\ \label{esplitrans}
\end{eqnarray}
where the vector $\vec{\xi}=(x,y)$ indicates the coordinates over
the target surface $S$, while $\vec{\kappa}=(\kappa_x,\kappa_y)$
and $\vec{\tau}=(k_x,k_y)$ indicate the transverse components of
the wave-vectors corresponding to the {\em pseudophotons\/} and
the transition radiation fields respectively.

As evident from the different role played in (\ref{esplitrans}) by
the Fourier transform of the transverse distribution functions
$\rho_u(k_u)$ $(u=x,y)$ with respect to the longitudinal one
$\rho_z(k_z)$, the three-dimensional generalization of
Eq.(\ref{coe}) by substituting the {\em longitudinal form
factor\/} with a three-dimensional one such as in Eq.(\ref{triff})
appears to be incorrect. It is also obvious to notice that
Eq.(\ref{esplitrans}) reduces to the well known {\em single
particle\/} transition radiation field \cite{termik} when $N=1$
and $\rho_u(k_u)=1$ $(u=x,y,z)$.

\subsection{Transition radiation spectrum:
from the discrete three-dimensional distribution to the continuous
limit}

In practical situations, as those occurring in an accelerator
machine, the number of the electrons composing a bunch is so large
that it can be reasonably described by means of a continuous
probability distribution function ({\em continuous density
limit\/}). Therefore, in order to derive the expression of the
actually detected transition radiation spectrum, an average
procedure has to be carried out:
\begin{equation}
\frac{d^2I}{d\Omega d\omega}=\left\langle%
\frac{d^2I_{dis}}{d\Omega d\omega}%
\right\rangle_{ens}\label{ens}
\end{equation}
where in the second term of the previous statement the energy
radiated - per solid angle and frequency units - by a particular
discrete charge distribution is averaged over the ensemble of the
infinite series of the equivalent discrete particle configurations
\cite{hirsch}.

Supposing $N$ electrons moving with a common uniform velocity
$\vec{w}$, the following expression
\begin{equation}
\rho(\vec{k})=\frac{1}{N}\sum_{j=1}^{N}e^{-i\vec{k}\cdot%
\vec{r}_j}
\end{equation}
is the Fourier transform of the spatial discrete distribution
\begin{eqnarray}
\rho(\vec{r},t)=\frac{1}{N}\sum_{j=1}^N\delta(\vec{r}-%
\vec{r}_j)\label{spdis}.
\end{eqnarray}
A useful notation, distinguishing in the Fourier transform of the
charge distribution the contribution
[$\tilde{\rho}_j(k_x,k_y)=e^{-i(k_xx_j+k_yy_j)}$] referred to
electrons belonging to the same slice (common $z$ coordinate)
\begin{equation}
\rho(\vec{k})=\frac{1}{N}\sum_{j=1}^{N}e^{-ik_zz_j}%
\tilde{\rho}_j(k_x,k_y)\label{fourslice},
\end{equation}
allows also to outline that spatial coherent effects are expected
to be observed in the spectrum, when the beam can be considered -
with respect to the observed wave-length - as a discrete sequence
of transverse continuous charge slices. In addition temporal
coherent effects are expected to arise in the spectrum when the
wave-length is so long that the slices sequence has to be
considered as a continuous charge distribution also in the
longitudinal direction.

As more extensively described in the Appendix, the complete formal
expression of the transition radiation spectrum - produced by a
three-dimensional electron beam colliding normally with a metallic
screen being at rest in the vacuum - can be expressed as
\begin{eqnarray}
\frac{d^2I(\vec{\tau},\omega)}{d\Omega d\omega}=%
\frac{d^2I^s(\vec{\tau},\omega)}{d\Omega
d\omega}[N+N(N-1)F(\omega)],\label{finalespec}
\end{eqnarray}
where
\begin{eqnarray}
&&\frac{d^2I^s(\vec{\tau},\omega)}{d\Omega d\omega}=%
\frac{c(ke)^2}{(2\pi)^4(\pi w)^2}\times\nonumber\\
&&\times\sum\limits_{u=x,y}\left|\int\limits_{S}d\vec{\xi}\,%
\int d\vec{\kappa}\,%
\frac{\kappa_u\rho_x(\kappa_x)\rho_y(\kappa_y)%
e^{i(\vec{\tau}-\vec{\kappa})\cdot\vec{\xi}}}%
{[\kappa_x^2+\kappa_y^2+\alpha^2]}\right|^2\nonumber\\
\label{fin-tot-spec}
\end{eqnarray}
under the hypothesis of the {\em far field approximation\/} and
$F(\omega)$ is the {\em longitudinal form factor\/}
[Eq.(\ref{ff})].

Some comments about the last statements. (1) When temporal
coherent conditions are occurring, the transition radiation
spectrum in the three-dimensional case shows the same dependence
on the {\em longitudinal form factor\/} [Eq.(\ref{ff})] as in the
one-dimensional case [Eq.(\ref{coe})]. (2) The results stated by
Eqs.(\ref{finalespec}, \ref{fin-tot-spec}) are consistent with the
ones valid for one-dimensional charged beam when
$\rho_x(\kappa_x)=\rho_y(\kappa_y)=1$ or when the observed
wave-length is longer than the transverse size of the beam. (3)
Conversely, when the observed wave-length is shorter than the
transverse size of the beam, {\em source size effects\/} appear
evident in the spectrum via a dependence on the Fourier transforms
of the transverse distribution of the beam
$(\rho_u(\kappa_u),u=x,y)$. Therefore, both the measurements of
the energy and of the transverse size of the beam are not
independent each other when the spectrum shows spatial incoherent
characteristics.

\section{Gaussian beam colliding with an infinitely
extended metallic screen}

In the ideal and handy case of a  three-dimensional gaussian beam
normally hitting a metallic screen $S$ of infinite size, the
analytical expression for the transition radiation spectrum can be
derived from Eqs.(\ref{fin-tot-spec}):
\begin{eqnarray}
&&\frac{d^2I^s(k_x,k_y,\omega)}{d\Omega d\omega}=%
\frac{c(ke)^2}{(\pi w)^2}\frac{(k_x^2+k_y^2)\left|\rho_x(k_x)%
\right|^2%
\left|\rho_y(k_y)\right|^2}{[k_x^2+k_y^2+\alpha^2]^2}.\nonumber\\
\label{spec-inf-scr}
\end{eqnarray}
It is easy to check that for a point-like charge
($\rho_x(k_x)=\rho_y(k_y)=1)$ Eq.(\ref{spec-inf-scr}) reduces to
the well known frequency independent {\em single particle
spectrum\/} \cite{termik,gari,bass}:
\begin{equation}
\frac{d^2I_e}{d\theta d\omega}=\frac{2(e\beta)^2}{c\pi}%
\frac{\sin^3 \theta}%
{(1-\beta^2\cos^2\theta)^2},\label{spec-inf-scr-esp}
\end{equation}
where $k_x^2+k_y^2=(k\sin\theta)^2$ and the polar angle $\theta$
is referred to the $z$ axis picked to be normal to the target
surface.

While for a three-dimensional bunch, having a gaussian transverse
distribution
$(\rho_u(k_u)=e^{-\frac{(k_u\sigma_u)^2}{2}},u=x,y,z)$, the
Eq.(\ref{spec-inf-scr}) can be written as
\begin{eqnarray}
\frac{d^2I^s}{d\Omega d\omega}&&=\frac{(e\beta)^2}{c\pi^2}%
\frac{\sin^2 (\theta) e^{-(k\sin\theta)^2(\sigma_x^2\cos^2\phi+%
\sigma_y^2\sin^2\phi)}}{(1-\beta^2\cos^2\theta)^2},%
\label{inf-scr-tras}
\end{eqnarray}
which, integrated over $(0,2\pi)$ with respect to the azimuthal
angle $\phi$, reduces to
\begin{eqnarray}
\frac{d^2I^s}{d\theta d\omega}=&&%
\frac{d^2I_e}{d\theta d\omega}%
e^{-\frac{(k\sin\theta)^2(\sigma_x^2+\sigma_y^2)}{2}}%
I_0[(k\sin\theta)^2\frac{(\sigma_x^2-\sigma_y^2)}{2}],\nonumber\\
\label{inf-scr-tras-asym}
\end{eqnarray}
where
$I_\nu(z)=e^{\frac{-i\pi\nu}{2}}J_\nu(e^{\frac{i\pi}{2}}z)\,%
(-\pi<argz\leq\pi/2)$
and $J_\nu(z)$ is the Bessel function of the first kind
\cite{grads}. In fig.1 the angular distribution
[Eq.(\ref{inf-scr-tras-asym})] corresponding to a cylindrical
symmetric gaussian beam $(\sigma=\sigma_x=\sigma_y)$ is compared
with the {\em single particle spectrum\/}
[Eq.(\ref{spec-inf-scr-esp})] for different values of the beam
energy $(E = 500 MeV, 5 GeV, 50 GeV)$ and of the
$\chi=\sigma/\lambda$ parameter, where $\lambda$ is the observed
wave-length.
%\begin{figure}
%\vspace{172mm}%
%\caption{Intensity distribution (A.U.) {\em vs\/} polar angle
%$\theta$ (rad) for several values of the beam energy (E=0.5,\,5
%and 50 GeV). The {\em single particle\/} distribution (a) is
%compared with the ones corresponding to a cylindrical symmetrical
%gaussian beam ($\sigma_x=\sigma_y=\sigma$) for different values of
%the $\chi=\sigma/\lambda$ parameter: $\chi=2\cdot10^2$ (b),
%$\chi=10^3$ (c), $\chi=2\cdot10^3$ (d). The value
%$\chi=2\cdot10^3$ corresponds to a transverse size $\sigma=1\,mm$
%and to an observed wave-length $\lambda=0.5\,\mu m$}
%\end{figure}

The shift of the polar angle distribution of the transition
radiation intensity - shown in fig.1 - is also meaningful and
instructive of the circumstance that beam energy measurements,
based on the analysis of the angular distribution of the spectrum,
can be - depending on the observed wave-length - strongly affected
by the beam transverse size at the low energy regime. In fact,
according to the {\em single particle\/} theory, the angular
position $\theta_{max}$ of the intensity peak is typically
considered as a rough estimate of the beam energy
$(\theta_{max}\approx mc^2/E)$.

Finally - with regard to Eq.(\ref{inf-scr-tras-asym}) - the
complete expression for the transition radiation spectrum
corresponding to a three dimensional gaussian beam passing through
an infinite metallic screen is
\begin{equation}
\frac{d^2I}{d\theta d\omega}=[N+N(N-1)F(\omega)]%
\frac{d^2I^s}{d\theta d\omega}\label{inf-scr-tot}
\end{equation}
with $F(\omega)=e^{-(\frac{\omega \sigma_z}{c})^2}$.

\section{Source size dependence of the spectrum and main
experimental consequences}

From Eqs.(\ref{spec-inf-scr}, \ref{inf-scr-tras}) it can be
observed that the spectral angular distribution carries
information about the transverse beam size. Then, as far as the
arguments treated in this paper can be considered valid, the
analysis of the angular distribution of the transition radiation
intensity offers information not only about the beam energy, but
also about the beam size. From such a circumstance several
consequences concerning beam diagnostics applications of the
transition radiation and deeper understanding of its phenomelogy
can be drawn.

First of all the beam energy measurement based on the usual {\em
single particle\/} analysis of the angular distribution of the
transition radiation in the optical region can be affected by a
systematic error as larger as longer is the transverse beam size
with respect to the observed wave-length. Moreover, such a
dependence of the spectrum on the Fourier transform of the
transverse charge distribution offers interesting and suggestive
opportunities in the field of the beam diagnostic applications of
the transition radiation. In fact, although the technique to
measure the transverse size of the beam by analyzing the angular
distribution of the intensity offers a poor appealing as beam
diagnostic tool in the optical region (OTR,
\cite{wartski,rule,rul-fiorito}), nevertheless it appears to be
promising in the infrared-micrometer analysis
\cite{shibata,orla,ru-fi} of those transition radiation spectra
produced by intense and high energy charged beams passing through
a hole in a metallic screen, which are properly referred as
diffraction radiation phenomena \cite{termik}.

Related to the understanding of the phenomelogy, it is to be
noticed that, according to Eq.(\ref{spec-inf-scr}), the spectral
angular distribution decreases to zero when the frequency tends to
infinity, and that such a behaviour is in agreement with the high
frequency features of the experimentally observed transition
radiation spectra \cite{ginz,jack}. Moreover, it is to be stressed
that such agreement - in this context necessarily qualitative -
with the experimental results does not follow from any assumption
on the frequency dependence of the dielectric constant of the
metallic screen. Conversely it is well known that, in order to
explain such a feature in the measured transition radiation
spectra, it is necessary to dismiss in the high frequency regime
the hypothesis of considering ideal conductor properties for the
metallic screen (dielectric constant $\epsilon_{cond}=\infty$) and
to take into account the real dielectric properties
$(\epsilon_{cond}\approx 1)$ when the frequency is higher than the
plasma frequency $\omega_p$ \cite{jack}, according to:
\begin{equation}
\epsilon_{cond}(\omega)=1-\left(\frac{\omega_p}{\omega}\right)^2\,.
\end{equation}
Furthermore, other and more qualified confirmations to the present
approach to the transition radiation theory for real electron
beams follow from a comparison with some other well-known features
of the experimental spectra. In fact, it is possible to derive the
analytical expression for the total energy emitted during the
radiative mechanism by integrating Eq.(\ref{inf-scr-tras-asym})
with respect to the polar angle $\theta$ and the wave-number $k$.
For such a purpose the transverse size of the beam is assumed so
large that the {\em ideal conductor hypothesis\/}
$(\epsilon_{cond}=\infty)$ can be considered valid in that part of
the frequency band effectively of interest in the integration.
Moreover, assuming for simplicity a gaussian cylindrical symmetry
$(\sigma_x=\sigma_y=\sigma)$ for an electron beam with energy $E$,
the total - ``forward'' and ``backward'' - energy emitted during
the collision can be expressed as:
\begin{eqnarray}
W^{tr}=\frac{\sqrt{\pi}(e\beta)^2}{2\sigma}%
\frac{E}{mc^2}\,.\label{tot-energy}
\end{eqnarray}
In the range of validity of the just referred hypothesis and
without introducing any frequency cut-off, the analytical formula
expressed by Eq.(\ref{tot-energy}) allows a plain explanation of
the experimental main features observed in the emitted transition
radiation energy in both the low and the high beam energy limits.
The previous statement is a particular case of the most general
expression valid for an asymmetric gaussian distribution
$(\sigma_x\neq\sigma_y)$:
\begin{eqnarray}
W^{tr}=\frac{\sqrt{\pi}(e\beta)^2}{2\sqrt{\sigma_x\sigma_y}}%
\frac{E}{mc^2}F(\frac{1}{4},\frac{1}{4};1;-\frac{(\sigma_x^2-\sigma_y^2)^2}%
{4\sigma_x^2\sigma_y^2})\label{tot-energy-asym}
\end{eqnarray}
where $F(\alpha,\beta;\gamma;z)$ is the Hypergeometric function
\cite{grads}.

As comment to the last result, it should be noticed that,
approaching the beam axis over the boundary surface, the {\em
pseudophotons\/} field [Eq.(\ref{esplipse})] diverges to infinity
in the case of a single particle, while it maintains a finite
value in the case of a three-dimensional charge distribution, as
expected for the electric field inside the core of a spatially
extended charge distribution. In fact, using spherical coordinates
and referring the polar angle $\theta$ to the $z$ axis normal to
the target surface lying on the $(x,y)$ plane, it can be
demonstrated that, for a gaussian beam with a cylindrical
symmetry, the {\em pseudophotons\/} electric field
[Eq.(\ref{esplipse})] over the boundary surface is
\begin{equation}
E^{ps}(\rho)\propto\left|\int%
d\kappa\,\frac{\kappa^2%
e^{-\frac{(\kappa\sigma)^2}{2}}J_1(\kappa\rho)}%
{\kappa^2+\alpha^2}\right|\label{radial-elec}
\end{equation}
where $\rho=\sqrt{x^2+y^2}$ and $\kappa=k\sin\theta$. It can be
easily verified that for a point-like charge $(\sigma=0)$ the
field Eq.(\ref{radial-elec}) diverges to infinity $(\sim 1/\rho)$
as $\rho\rightarrow 0$ , while in the case of a transverse
gaussian beam [Eq.(\ref{radial-elec})] continues to maintain a
finite and bounded value $(\sim 1/\sigma)$.

\section{Conclusions}

In the framework of the {\em pseudophotons method\/} and under the
hypothesis of the {\em far field approximation\/} it has been
demonstrated that, in the case of a three-dimensional charged beam
colliding with a metallic target, the temporal coherent part of
the spectrum depends on the {\em longitudinal form factor\/} in
the same manner as in the one-dimensional case. It has been also
demonstrated that the transverse beam size affects the spectrum
via the Fourier transform of the transverse charge distribution
and that the spectrum becomes insensitive to it - reducing to the
well known {\em single particle spectrum\/} - when the observed
wave-length is larger than the transverse beam size. As reasonable
to expect, the {\em single particle spectrum\/} can be interpreted
as the spatial coherent limit of the transition radiation spectrum
corresponding to a three-dimensional beam. The demonstrated
evidence in the spectrum of the spatial incoherent effects due to
the transverse size of the beam - on the one hand - allows a
better comprehension of some aspects of the transition radiation
phenomenology - on the other hand - offers new perspectives in the
use of the transition radiation as beam diagnostic tool.

In fact, as the transverse size of the beam is increasing, the
high frequency damping feature observed in the experimental
transition radiation spectra should depend more on the {\em source
size\/} effects than on the frequency dependence of the dielectric
constant of the metallic screen. Moreover, supposing as dominant
such a mechanism, it is also possible to derive an analytical
formula for the total radiated energy showing an unexpected
dependence on the reciprocal of the transverse beam size besides
the expected one on both the square of the charge velocity and
energy.

From the point of view of the beam diagnostic implications, the
dependence on {\em transverse form factor\/} appears to be an
useful tool to derive information about the transverse beam size
from the analysis of the spectral angular distribution, even
observing the radiation in conditions of temporal incoherence and
in non-optical wave-length regions, when the radiative mechanism
is produced in a non-destructive way by a charged beam passing
through a slit in a metallic target (diffraction radiation). In
fact from Eqs.(\ref{inf-scr-tras-asym}), supposing as known the
beam energy, it appears in principle possible to measure the beam
size by observing the radiation in conditions of spatial
incoherence, i.e., by selecting - by means of a monochromator -
from the spectrum a wave-length much shorter than the transverse
size itself and by analyzing the corresponding angular
distribution.

\section*{Acknowledgement}

This work was partially developed after the accomplishment of the
Ph.D. in Physics at the "Tor Vergata" University of Rome during a
period of hospitality at Laboratori Nazionali di Frascati of INFN,
where I took advantage of useful discussions with Dr. M.
Castellano, Prof. S. Tazzari, Dr. F. Tazzioli and Dr. V. Verzilov.
For encouragement, precious comments and helpful discussions I
would like to express my acknowledgments to Dr. A. Aiello, Prof.
C. Bernardini, Dr. C. Curceanu (Petrascu), Dr. G. Dattoli, Dr. S.
Frullani, Prof. L. Paoluzi, Dr. L. Picardi and Prof. S. Segre.

\appendix
\section{Appendix}

The formal expression of the transition radiation spectrum
produced by a three-dimensional charged beam
[Eqs.(\ref{finalespec}, \ref{fin-tot-spec})] is obtained supposing
as infinite the number of the electrons composing the bunch.
According to the {\em continuous density limit\/} \cite{hirsch}
one can assume: (1) The ensemble average of Eq.(\ref{ens}) does
not depend on $N$. (2) All the sum operations over $N$ can be
extended to infinity and, thus, substituted with integral
operations, while the charge discrete spatial distribution
[Eq.(\ref{spdis})] can be identified with the corresponding mean
continuous probability distribution function
[Eq.(\ref{con-sp-dis})].

Disregarding the common phase factor and indicating the integrand
in Eq.(\ref{esplitrans}) as
\begin{eqnarray}
H_{x,y}(\vec{\tau},\vec{\kappa},\vec{\xi})=&&%
\frac{kNe}{2\pi^2Rw}\times\nonumber\\
&&\times\frac{\kappa_{x,y}e^{i[(k_x-\kappa_x)x+(k_y-\kappa_y)y]}}%
{\epsilon[\kappa_x^2+\kappa_y^2+\alpha^2]},\label{acca}
\end{eqnarray}
and according to Eq.(\ref{fourslice}), the energy radiated - per
solid angle and frequency units - by a particular discrete charge
distribution can be expressed as:
\begin{eqnarray}
\frac{d^2I_{dis}(\vec{\tau})}{d\Omega d\omega}=&&
\frac{cR^2}{4\pi^2N^2}\times\nonumber\\
&&\times\sum\limits_{u=x,y}[N\left|S_u(\vec{\tau})\right|^2+%
N(N-1)\left|T_u(\vec{\tau})\right|^2],\nonumber\\
\label{esplitdis}
\end{eqnarray}
where
\begin{eqnarray}
\left|S_u(\vec{\tau})\right|^2=&&%
\int d\vec{\xi}d\vec{\kappa}%
d\vec{\xi}'d\vec{\kappa}'\, H_u(\vec{\tau},%
\vec{\kappa},\vec{\xi})H_u^{*}(\vec{\tau},%
\vec{\kappa}',\vec{\xi}')\times\nonumber\\
&&\times\sum\limits_{j=1}^{N}\frac{\tilde{\rho}_j(\vec{\kappa})%
\tilde{\rho}_j^{*}(\vec{\kappa}')}{N} \label{dis-spa}
\end{eqnarray}
and
\begin{eqnarray}
\left|T_u(\vec{\tau})\right|^2=&&%
\int d\vec{\xi}d\vec{\kappa}%
d\vec{\xi}'d\vec{\kappa}'\, H_u(\vec{\tau},%
\vec{\kappa},\vec{\xi})H_u^{*}(\vec{\tau},%
\vec{\kappa}',\vec{\xi}')\times\nonumber\\
&&\times\sum\limits_{j,m(j\neq m)=1}^{N}%
\frac{e^{i(\omega/w)(z_m-z_j)}%
\tilde{\rho}_j(\vec{\kappa})\tilde{\rho}_m^{*}%
(\vec{\kappa}')}{N(N-1)}\nonumber\\ \label{dis-spa-tem}
\end{eqnarray}

The average over the ensemble of the discrete configurations of
the first term in Eq.(\ref{esplitdis}) is then proportional to
\begin{eqnarray}
\left\langle\left|S_u(\vec{\tau})\right|^2\right\rangle_{ens}%
=\left|\int d\vec{\xi}d\vec{\kappa}\,%
H_u(\vec{\tau},\vec{\kappa},\vec{\xi})%
\tilde{\rho}(\vec{\kappa})\right|^2\label{inco-con}
\end{eqnarray}
where
$\tilde{\rho}(\vec{\kappa})=\rho_x(\kappa_x)\rho_y(\kappa_y)$ is
the Fourier transform of the continuous probability distribution
$\rho_x(x)\rho_y(y)$ in the transverse plane. While the temporal
coherent contribution to the transition radiation spectrum is
\begin{eqnarray}
\left\langle\left|T_u(\vec{\tau})\right|^2\right\rangle_{ens}=%
F(\omega)\left|\int d\vec{\xi}d\vec{\kappa}\,%
H_u(\vec{\tau},\vec{\kappa},\vec{\xi})%
\tilde{\rho}(\vec{\kappa})\right|^2\nonumber\\ \label{fin_con-coe}
\end{eqnarray}
where $F(\omega)$ is the {\em longitudinal form factor\/}
[Eq.(\ref{ff})] under the hypothesis of an ultra-relativistic
electron beam traveling in the vacuum $(\epsilon=1, w\approx c)$.

The explicit expression of transition radiation spectrum stated by
Eqs.(\ref{finalespec}, \ref{fin-tot-spec}) follows from
Eqs.(\ref{inco-con},\ref{fin_con-coe}) according to
Eq.(\ref{acca}).

\newpage
\begin{center}
{\Large \bf Figure Captions }
\end{center}
FIGURE 1. Intensity distribution (A.U.) {\em vs\/} polar angle
$\theta$ (rad) for several values of the beam energy (E=0.5,\,5
and 50 GeV). The {\em single particle\/} distribution (a) is
compared with the ones corresponding to a cylindrical symmetrical
gaussian beam ($\sigma_x=\sigma_y=\sigma$) for different values of
the $\chi=\sigma/\lambda$ parameter: $\chi=2\cdot10^2$ (b),
$\chi=10^3$ (c), $\chi=2\cdot10^3$ (d). The value
$\chi=2\cdot10^3$ corresponds to a transverse size $\sigma=1\,mm$
and to an observed wave-length $\lambda=0.5\,\mu m$
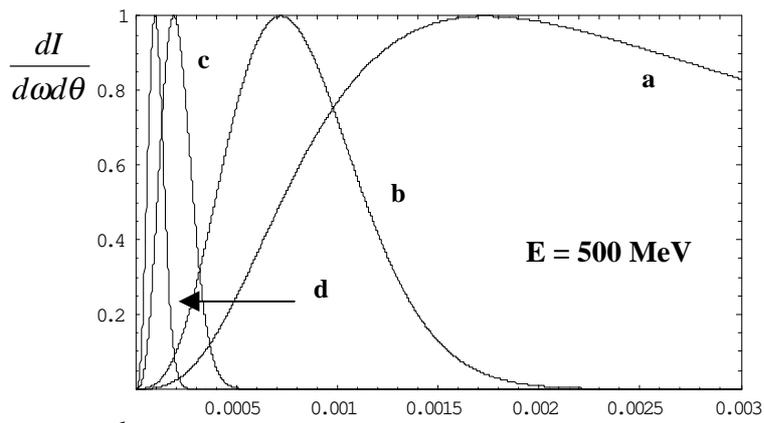
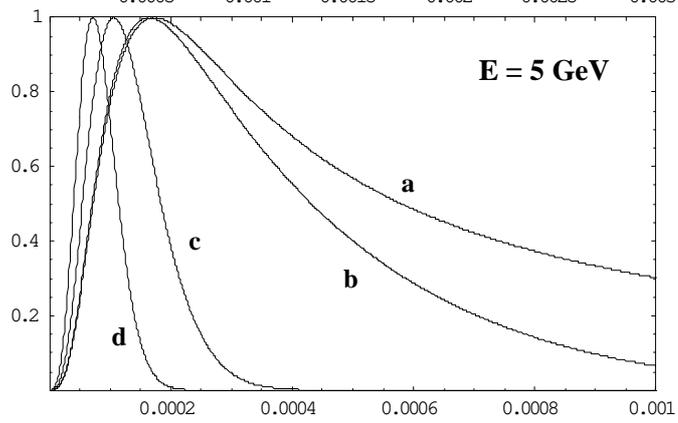
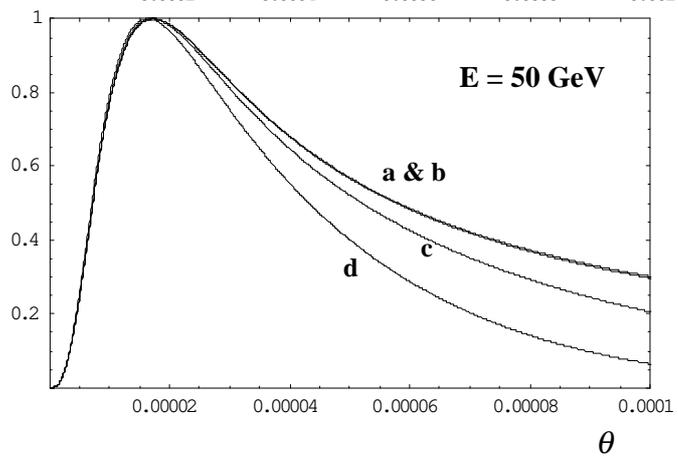
\end{document}